\documentclass[%
 reprint,
showpacs,preprintnumbers,
 amsmath,amssymb,
 aps,
prl,
floatfix,
]{revtex4-1}

\usepackage{graphicx}% Include figure files
\usepackage{dcolumn}% Align table columns on decimal point
\usepackage{bm}% bold math

\begin{document}

%\preprint{APS/123-QED}

\title{Spontaneous jamming and unjamming in a hopper with multiple
  exit orifices}

\author{Amit Kunte} \affiliation{Chemical Engineering Division,
  National Chemical Laboratory, Pune 411008 India} \author{Pankaj
  Doshi} \email{p.doshi@ncl.res.in} \affiliation{Chemical Engineering
  Division, National Chemical Laboratory, Pune 411008 India}
\author{Ashish V. Orpe} \email{av.orpe@ncl.res.in}
\affiliation{Chemical Engineering Division, National Chemical
  Laboratory, Pune 411008 India}

\date{\today}% It is always \today, today,
% but any date may be explicitly specified

\begin{abstract}
  We show that the flow of granular material inside a 2-dimensional
  flat bottomed hopper is altered significantly by having more than
  one exit orifice.  For the hoppers with small orifice widths,
  intermittent flow through one orifice enables the resumption of flow
  through the adjacent jammed orifice, thus displaying a sequence of
  jamming and unjamming events.  Using discrete element simulations,
  we show that the total amount of granular material (i.e. avalanche
  size) emanating from all the orifices combined can be enhanced by
  about an order of magnitude difference by simply adjusting the
  inter-orifice distance. The unjamming is driven primarily by
  fluctuations alone when the inter-orifice distance is large, but
  when the orifices are brought close enough, the fluctuations
  along with the mean flow cause the flow unjamming.
\end{abstract}

\pacs{45.70.Mg,47.57.Gc}% PACS, the Physics and Astronomy
% Classification Scheme.
% \keywords{Suggested keywords}%Use showkeys class option if keyword
% display desired
\maketitle

An assembly of discrete, non-cohesive particles, aka dry granular
media, while trying to flow through a narrow opening can clog or jam,
an occurrence widely observed in particle drainage through silos and
hoppers which are used ubiquitously in several industrial
applications.  The jamming at the orifice is caused due to few
particles forming a stable arch at the exit~\cite{to01,tang11}
eventually causing the entire system to halt abruptly. The particles
constituting the arch can be from anywhere in the system~\cite{tang11}
and while occurrence of the arch is quite unpredictable, some
information can be obtained through the spatial distribution of the
velocity fluctuations in the system~\cite{tewari13}. The shape of the
arch, though, can be predicted quite well using a random walk
model~\cite{to01}.

For a three dimensional hopper there exists a critical orifice width
above which the system never jams~\cite{zuriguel03} while no such
limit exists for a two dimensional system which can jam for large
enough orifice widths~\cite{janda08}. It has been shown that the
output flow rate from a hopper can be increased as large as $10$\% by
placing suitable inserts within at appropriate locations
~\cite{yang01,zelinski09,alonso12}. This also decreases the
probability of jamming by about two orders of
magnitude~\cite{zuriguel11}. These studies have shown that the
reduction in the tendency to jam is due to the reduction of the
pressure in the region of arch formation. The motivation for placing
inserts is derived from parallel interesting studies carried out to
alter the flow behavior of pedestrians from a crowded
room~\cite{frank11,yanagisawa09}.

Over here we report an interesting observation about the jamming
behavior in a 2-dimensional hopper having two exit orifices of same
width placed far away from each other. Whenever one of the orifice
jams, which is expected given the orifice width incorporated, it
unjams spontaneously if the flow is occurring through other orifice.
Effectively, flow continues to
occur through both orifices, alternately or together, for much longer
duration than expected for a hopper with single orifice.  The overall
duration of flow, consequently the tendency of jamming, can be altered
simply by changing the inter orifice distance.  Such non-local
interaction between regions exhibiting differing flow behavior has
been observed previously, but in different
configurations~\cite{nichol10,reddy11} and under different flow
conditions. In these studies it was shown that the origin of such
non-local interaction can be attributed to a self activated process
within which the stress fluctuations induced by a localized shear
causes the material to yield and flow elsewhere~\cite{reddy11}.

\begin{figure}[b]
  \includegraphics[width=0.9\linewidth]{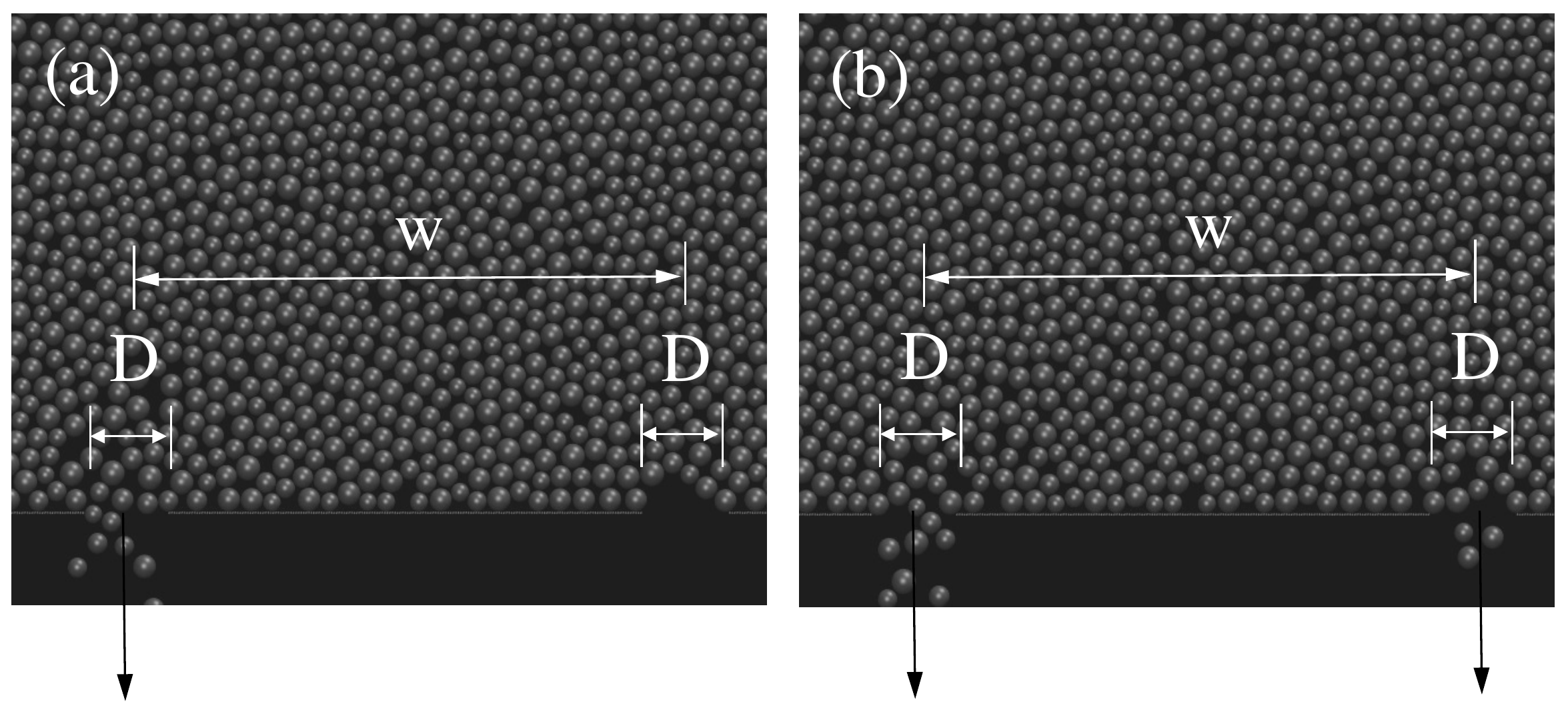}
  \caption{Sample snapshots of the flow occurrence in a two-orifice
    hopper for a specified $w$ and $D$. (a) Flow occurring through the
    left orifice while the right orifice is jammed. (b) Spontaneous
    flow re-initialization through the right orifice at a later time.}
  \label{fig1}
\end{figure}

In the present system, we conjecture that the observed behavior of
spontaneous unjamming occurs due to rearrangement of the particles in
the jammed region above one orifice. This unjamming behavior can be
correlated with the fluctuations induced in the system due to flow
from the nearby orifice. We believe that such non-local interaction
between two widely separate regions can be used to systematically
alter the jamming behavior in a hopper non-intrusively which could be
quite significant for several industrial operations. We have explored
this behavior in great detail through DEM simulations of soft
particles. We measure the mean avalanche size for varying conditions
and show their correlation with the fluctuations measured as root mean
squared (r.m.s.) values.

The DEM simulations are carried out using the Large Atomic/Molecular
Massively Parallel Simulator (LAMMPS) developed at Sandia National
Laboratories~\cite{plimp95,lammps}. The simulation employs Hookean
force between two contacting particles described in detail
elsewhere~\cite{rycroft09}. All the simulation parameters are the same
as used in the systematic study of hopper flows carried out
previously~\cite{rycroft09} except for a higher normal elastic
constant ($k_n = 2 \times 10^6 mgd$) which corresponds to a more
stiffer particle. The inter-particle friction coefficient ($\mu$) is
varied from $0.2$ to $0.8$ with no qualitative difference between the
results. Over here, we report the results obtained for $\mu = 0.5$.

The simulation geometry consists of a 2-dimensional hopper of height
about $1.2$ times its width and thickness $1$d, where $d$ is the mean
particle diameter with a polydispersity of $15$ \%. The sidewalls are
created out of the largest particles (rough wall) while the bottom
surface is created from the smallest particles (relatively smooth
wall) by freezing the particles so that their translational and
angular velocities are kept zero throughout the simulation.  The
hopper has two orifices of width $D$ placed at an inter-orifice
distance
$w$, both defined in terms of mean particle diameter $d$. The hopper
width is large enough to prevent any confinement effects due to their
proximity to either orifice. The fill height is maintained
constant by re-positioning the particles exiting from the orifice just
above the free surface.  The hopper is filled using the sedimentation
method as suggested previously~\cite{landry03} in which a dilute
packing of non-overlapping particles is created in a simulation box
and allowed to settle under the influence of gravity. The simulation
is run for a significant time so that the kinetic energy per particle
is less than $10^{-8} mgd$ resulting into a quiescent packing of
desired fill height $H$ in the hopper.

The flow through the hopper is initiated by opening both the orifices
simultaneously. The orifice width is chosen to be small enough to
cause jamming after certain period of flow. After the flow is
initialized, either of the orifice gets jammed but unjams again
spontaneously. Note that this unjamming would not have been possible
in the absence of second orifice through which the flow occurs for a
very short duration before jamming itself. The jamming-unjamming
sequence can, thus, flip from one orifice to other. Effectively the
flow occurs through either one or both the orifices at any given
time. After few of these jamming-unjamming sequences, the flowing
orifice jams before it can unjam the other orifice and the overall
flow stops. It is quite evident that the particles above the flowing
orifice transmit some information to the jammed orifice causing it to
unjam again. However, this information is available only for a short
duration before the flowing orifice jams on its own. After both
orifices jam, the flow is re-initiated by removing $2-3$ particles
from each of the arch. This procedure is continued to get significant
number of jamming events.  The total number of particles flowing out
from the hopper from the instant both orifices are opened until both
are jammed is defined as an avalanche size ($s$). The value of $s$
is found to depend on the values of $D$ and $w$ which we discuss
next. A sample snapshot of particles flowing through one orifice while
other being jammed is shown in Fig.~\ref{fig1}a. A small time later,
the jammed orifice starts to flow spontaneously as shown in
Fig.~\ref{fig1}b.

\begin{figure}
  \includegraphics[width=0.7\linewidth]{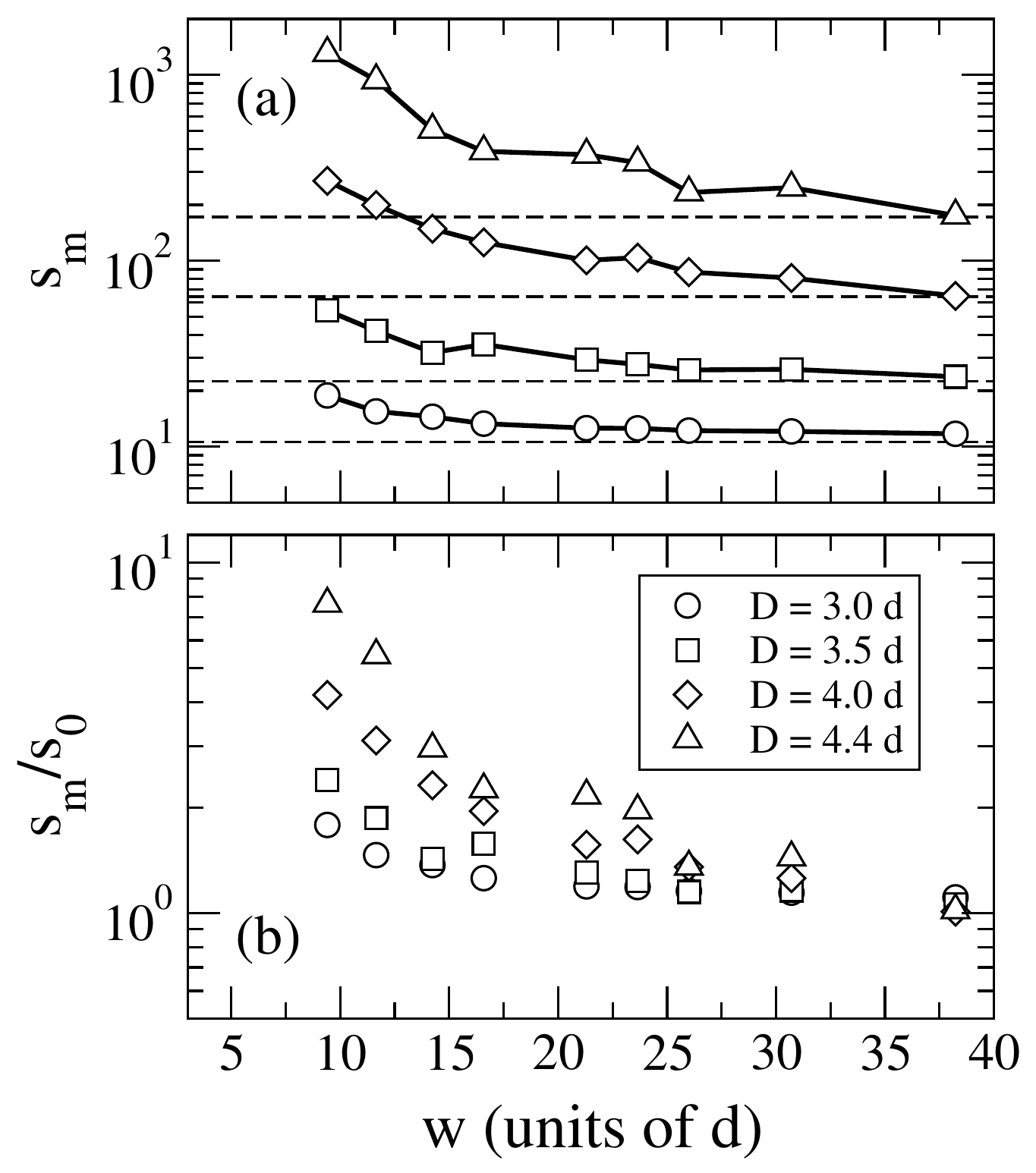}
  \caption{Jamming characteristics for varying distances ($w$) between
    orifices of different widths ($D$). (a) Mean avalanche size per
    unit orifice ($s_{m}$) plotted against the inter-orifice distance
    ($w$). Dashed lines corresponds to the avalanche size
    ($s_{0}$) for a hopper having single orifice of width $D$. (b)
    Scaled mean avalanche size ($s_{m}/s_{0}$) plotted against the
    inter-orifice distance ($w$).}
  \label{fig2}
\end{figure}

\begin{figure}
  \includegraphics[width=1.0\linewidth]{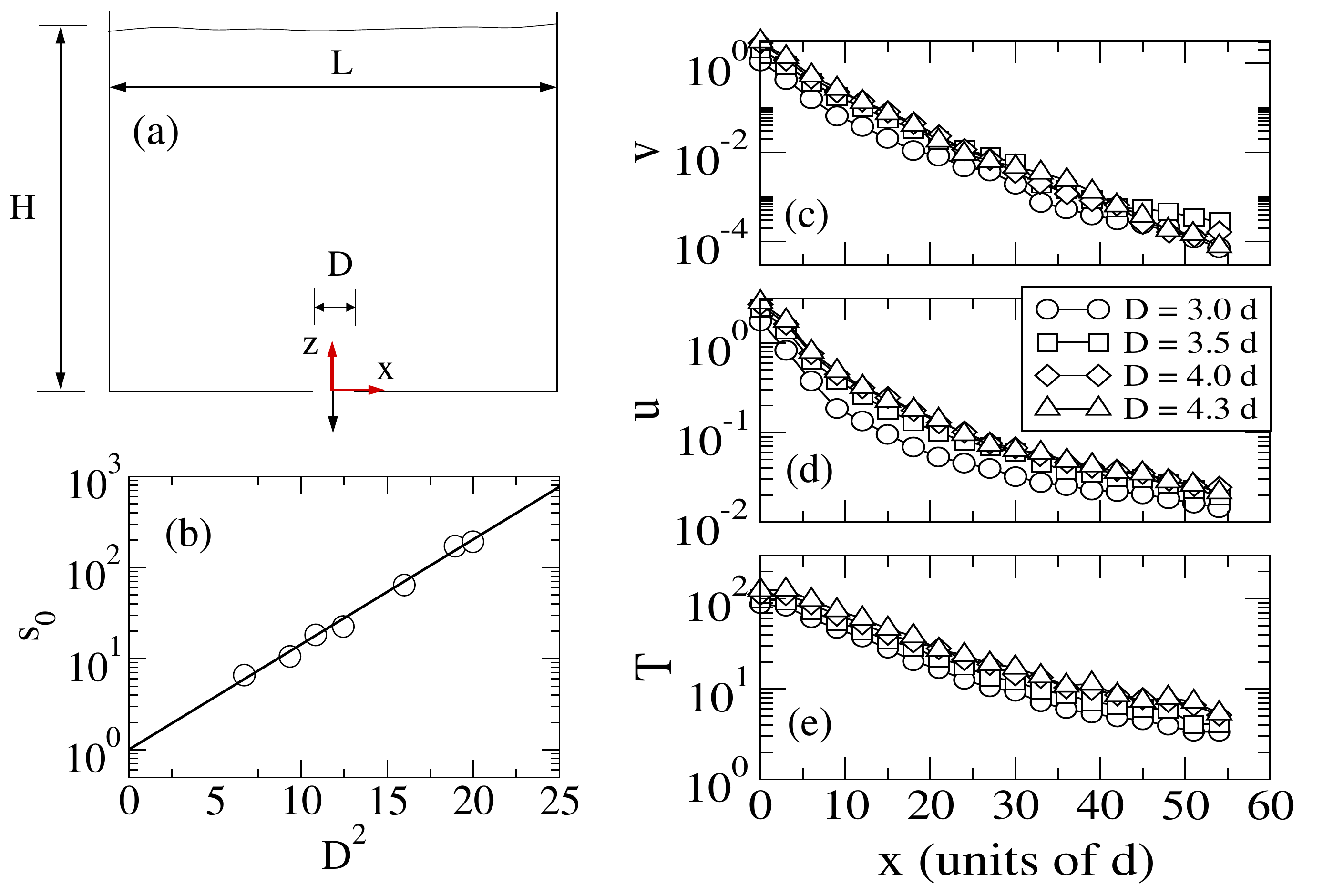}
  \caption{(a) Schematic of a single orifice hopper of width $L$ and
    height $H$. (b) Variation of mean avalanche size ($s_{0}$) for a
    hopper with single orifice of width $D$.  Variation of (c) mean
    velocity $v$, (d) r.m.s. velocity $u$ and (e) r.m.s. normal stress
    $T$ along the horizontal direction ($x$) obtained over a region
    $3d$ in $z-$direction for four different orifice widths. Please
    refer to text for the method to obtain mean and rms
    quantities. Mean velocity and rms velocity are measured in units
    of $d/\tau$, while rms stress is measured in units of $mg/d^{2}$.}
  \label{fig3}
\end{figure}

The distribution of avalanche sizes ($s$) for a given $D$ and $w$,
normalized by the mean avalanche size $\langle s \rangle$, exhibits an
exponential tail for all cases which is typical of the random nature
of the discrete avalanche events~\cite{zuriguel11}.  Fig.~\ref{fig2}a
shows the mean avalanche size per orifice ($s_{m} = \langle s
\rangle/2$) obtained for four different values of $D$ and several
inter-orifice distance ($w$). The behavior is qualitatively similar
for all the orifice sizes. The avalanche size ($s_{m}$) decreases
monotonically with increasing $w$ and at infinitely large $w$ it
asymptotically approaches a constant value, which corresponds to that
for a single orifice hopper (see Fig.~\ref{fig3}b). In this scenario,
the two orifices will function unaware of existence of the other and
no information is exchanged between the respective flowing
regions. For $w$ larger than those shown in Fig.\ref{fig2}a (but not
studied), an unjamming event can happen, but with much lesser
probability.  Now with decreasing $w$, the value of $s_m$ increases
and it grows quite rapidly for small enough $w$. This is the
consequence of the flow from one orifice aiding that through the other
in some way, thus, effectively increasing the total time period over
which flow occurs, hence larger $s_m$. In the limit the two orifices
are very close to each other ($w$ of the order of $D$ or lower), the
flow now occurs as if through a wider orifice ($\sim 2 D$).
Eventually it approaches the
asymptotic limit for a single orifice of width $2D$.  Within the
entire range of $w$, the value of $s_m$ is observed to vary by almost
an order of magnitude difference for the largest orifice width
considered. The mean avalanche size ($s_{m}$) normalized by the single
orifice avalanche size ($s_{0}$) is shown in Fig.~\ref{fig2}b. The
data shows reasonable collapse for $w > 20d$, while increasing scatter
is observed with decreasing $w$, which, perhaps, indicates a
non-linear dependence on the orifice width ($D$) and interacting flow
fields which cannot be scaled out by simple normalizing.

To elucidate the origin of the enhanced avalanche sizes, we perform
simulations in a hopper fitted with single orifice for different $D$
(see Fig.~\ref{fig3}a). The width of the hopper is more than twice the
maximum distance ($w$) used in two orifice system. For every orifice
size, several avalanches are obtained by re-initiating the flow post
jamming.  The mean avalanche size shows an exponential-squared
dependence on the orifice size ($e^{[0.26 D^2]}$) as shown in
Fig.~\ref{fig3}b and is in accordance with experimental results
obtained previously for a 2-dimensional system~\cite{janda08}
suggesting of the absence of a critical orifice width ($D$) separating
the flowing and jamming conditions.
 
We next obtain the profiles of mean velocity, velocity fluctuations
and normal stress fluctuations of the particles along the
$x-$direction. To calculate these quantities, we save snapshots of
particle positions at intervals of $0.0025 \tau$, where $\tau
=\sqrt{d/g}$, $g$ is the gravitational acceleration and integration
time step used in the simulations is $\delta t = 2.5 \times 10^{-5}
\tau$. The mean velocity is defined as $v(x) = \sqrt{\langle c_x
  \rangle^{2}+\langle c_z \rangle^{2}}$, while the average velocity
fluctuations are measured in terms of root mean squared (rms) velocity
defined as $u(x) =\sqrt{[\langle c_{x}^{2} \rangle - \langle c_{x}
  \rangle ^{2}]+[\langle c_{z}^{2} \rangle - \langle c_{z} \rangle
  ^{2}]}$. Here, $c_x$ and $c_z$ are the instantaneous horizontal and
vertical velocity components of every particle obtained from the
displacements between two successive snapshots. The simulation
algorithm also outputs the horizontal ($\sigma_{x}$) and vertical
($\sigma_{z}$) component of computed normal stress on each particle
within each snapshot. The normal stress fluctuations are captured in
terms of r.m.s. stress defined as $T(x) =\sqrt{[\langle \sigma_{x}^{2}
  \rangle - \langle \sigma_{x} \rangle ^{2}]+[\langle \sigma_{z}^{2}
  \rangle - \langle \sigma_{z} \rangle ^{2}]}$. In all the above
calculations, $\langle \rangle$ represents a temporal average over
several time instants of flow and spatial average over a $3d$ region
in $z-$direction located about $4d$ above the orifice. This spatial
region is chosen so as to capture the essential dynamics in a simple
manner. All the three quantities are shown in Fig.~\ref{fig3}c,d,e for
the four orifice widths. The profiles are symmetric
about the orifice and hence only the right half of the profiles are
shown in each case.

The magnitude of all the quantities increase with an increase in the
orifice width which is expected given the faster flow and
consequently, more collisions between particles. The profiles show
similar behavior for different orifice sizes and collapse quite nicely
(not shown) when normalized by the respective quantity at the $x = 0$.
The mean velocity decays much rapidly with distance from the orifice
while the r.m.s. velocity and stress show a much gradual decay. The
ratio of r.m.s. to the mean velocity increases progressively away from
the orifice and reaches a value of about $100$ by $x = 45d$. Almost
similar numbers are obtained for all the orifice sizes studied.

Now consider the scenario of a two orifice hopper with one of the
orifice located at $x = 0$ which is in unjammed (flowing)
state. Provided the other orifice located at some distance ($w$ or
$x$) is in a jammed state, the flow profile due to this orifice will
the same as shown in Fig.~\ref{fig3}c-e. The occurrence of flow
through the unjammed orifice over a time period (which corresponds to
the time interval between jamming and unjamming instances of second
orifice), causes particle rearrangements in the system and breaks the
arch above second orifice leading to flow. For an inter-orifice
distance lesser than $10d$, with the mean and r.m.s. velocity
magnitudes being of same order it is difficult to clearly isolate the
effect of each on the spontaneous unjamming behavior. However, for
larger inter-orifice distances ($> 20d$), the mean velocity is over
an order of magnitude lower than the rms velocity and the
fluctuations are expected to dominate the spontaneous unjamming
behavior. It is expected that the fluctuations too will decay at
infinitely large distances and will not able to re-initiate flow in
the jammed orifice. The flow and jamming behavior of the two orifices
in that case resemble that from the two isolated orifices unaware of
the other's existence. The mean avalanche size then approaches
that for a single orifice (dashed lines in Fig.~\ref{fig2}b).
 
\begin{figure}
  \includegraphics[width=1.0\linewidth]{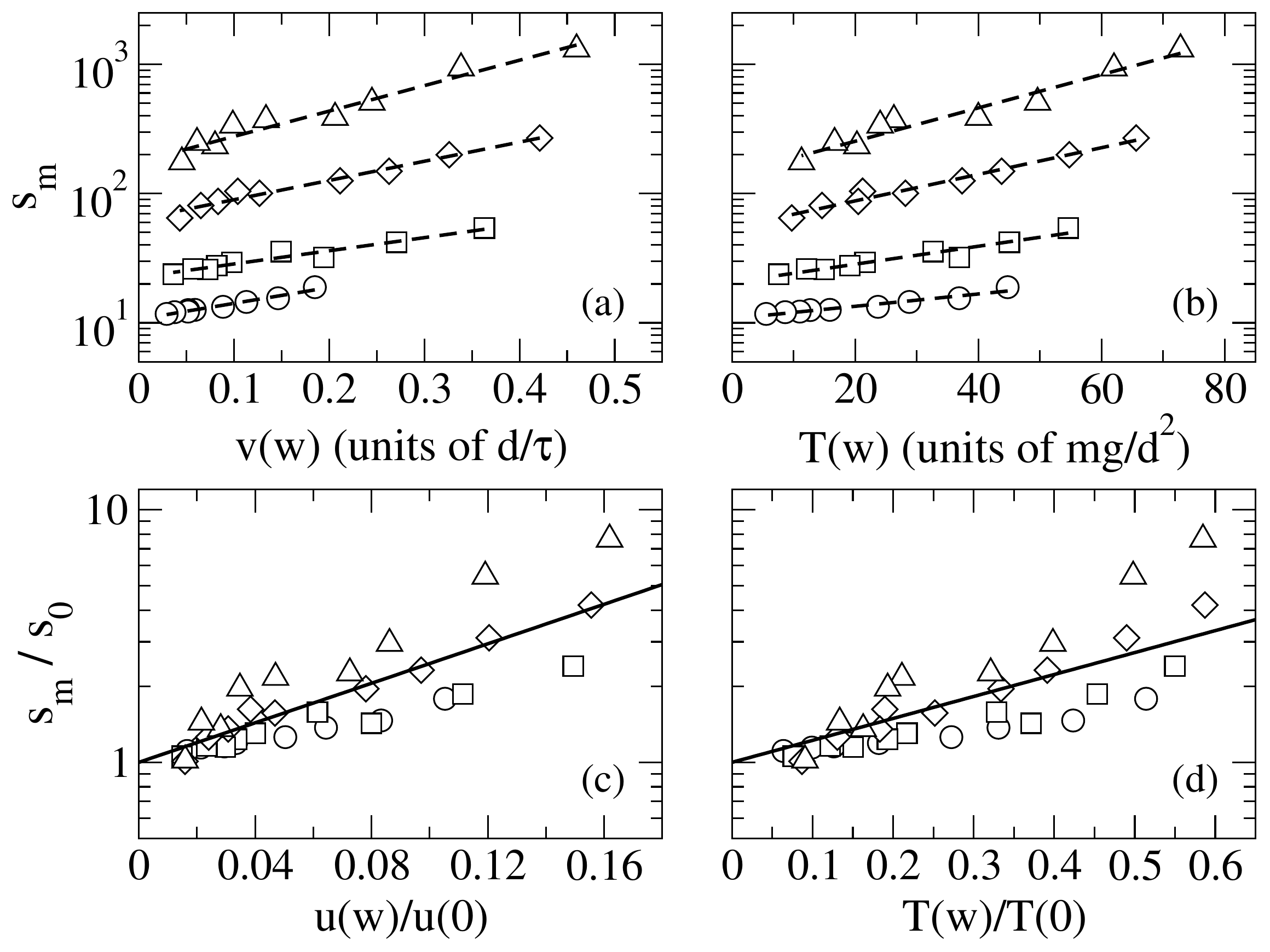}
  \caption{Correlation between mean avalanche size per unit orifice
    width ($s_{m}$) and (a) the r.m.s. velocity and (b) the r.m.s
    normal stress. All quantities obtained at distance ($w$).  Dashed
    lines show an exponential fit. (c, d) Data in (a) and (b)
    respectively after normalizing $s_{m}$ with mean avalanche size
    ($s_{0}$) for a hopper with single orifice. The solid line in (c)
    and (d) shows an exponential fit.}
  \label{fig4}
\end{figure}

Given the observed dependence of the mean avalanche size (see
Fig.~\ref{fig2}a) and that of the flow variables (see
Fig.~\ref{fig3}c,d,e) on the inter-orifice distance, we now proceed to
determine the relation between these quantities. We consider data only
for inter-orifice distance greater than $10d$ beyond which the mean
flow decreases rapidly compared to fluctuations.  Fig.~\ref{fig4}a and
b shows the avalanche sizes ($s_{m}$) for different inter-orifice
distances ($w$) plotted against the rms velocity $u(w)$ and rms normal
stress $T(w)$, respectively. All quantities are evaluated at the same
distance ($w$ or $x$) in all the cases. An exponential relation (shown
as dashed lines) is observed for all the cases where with the
y-intercept approximately equal to the avalanche size ($s_{0}$) for a
single orifice hopper (i.e. two orifices infinitely far away from each
other).  The normalized mean avalanche size ($s_{m}/s_{0}$) plotted
against the normalized r.m.s velocity and normalized r.m.s. normal
stress values is shown in Fig.~\ref{fig4}c and d, respectively.  The
rms velocity and normal stress are normalized with their corresponding
values at $x = 0$. The collapse is much better than that observed in
Fig.~\ref{fig2}b which indicates that the fluctuations do play a role
and $w$ is not the only parameter influencing the flow behavior. The
solid line shows an exponential fit.  For smaller $u(w)/u(0)$ and
$T(w)/T(0)$ ($w > 20d$), where the primary quantity responsible for
unjamming is fluctuations, the data collapse is quite good. However,
the scaling shows increasing scatter at increasing values of
$u(w)/u(0)$ and $T(w)/T(0)$ (or smaller $w$) which perhaps is
indicative of the more complex dependence on the mean flow in addition
to fluctuations and the inter-orifice width.  Similar scaling behavior
for both, r.m.s. velocity and r.m.s. normal stress data, is not quite
surprising as both represent average fluctuations in the system and
are quite inter-related to each other. Fluctuation of velocity is
expected to generate collisions between particles leading to
fluctuations in stresses and vice-versa.

In conclusion, our study suggest that the jamming occurrences within a
hopper can be altered non-intrusively using multiple orifices by
varying the inter-orifice distance. The hopper can be made to flow for
prolonged duration for a small enough orifice size by having another
orifice of same size at different distances. We observe that the
fluctuations arising locally cause rearrangements of particles in the
region located far away which leads to spontaneous unjamming. We would
also like to note that while the results have been reported for a
strictly 2-dimensional system, similar qualitative behavior is
observed in simulations of 3-dimensional or quasi-3d systems.  This
mechanism of spontaneous jamming-unjamming can be of immense
importance for (a) more detailed exploration of jamming
characteristics for granular systems studied
previously~\cite{bi11,liu10,corwin05,majmu05} and (b) investigating
other disordered systems which exhibit jamming, viz. bubbles escaping
from an orifice~\cite{bertho06} or colloidal hard spheres flowing
through a constriction~\cite{isa06}.  As a system of practical
utility, the multi-orifice hopper can be operated as an efficient
mixing device~\cite{kamath14} wherein the mixing can be initialized
through interaction between different zones of an hopper through
either random or controlled closing and opening of orifice.

\begin{acknowledgements}
  We thank Ken Kamrin and Chris Rycroft for several
  stimulating discussions, Mayuresh Kulkarni and Sandesh Kamath for
  help with some simulations and preliminary experiments and the
  funding from Department of Science and Technology, India, Grants
  No. SR/S3/CE/037/2009 and SR/S3/CE/0044/2010.
\end{acknowledgements}

\bibliography{jam}

\end{document}